\documentclass{appolb}
\usepackage{epsfig}

\begin{document}
\title{Dissipative superfluids, from cold atoms to quark matter%
\thanks{Presented at  Three Days of Strong Interactions, Wroclaw (Poland)    9. - 11. VII. 2009}%
}
\author{Massimo Mannarelli
\address{Instituto de Ciencias del Espacio (IEEC/CSIC) 
Campus Universitat Aut\` onoma de Barcelona,
Facultat de Ci\` encies, Torre C5 
E-08193 Bellaterra (Barcelona), Spain}
\and
Cristina Manuel
\address{ Instituto de Ciencias del Espacio (IEEC/CSIC) 
Campus Universitat Aut\` onoma de Barcelona,
Facultat de Ci\` encies, Torre C5 
E-08193 Bellaterra (Barcelona), Spain}
}
\maketitle
\begin{abstract}
Some results about dissipative processes in superfluids are presented. We focus on  fermionic superfluidity and restrict  our analysis to the contribution of phonons to  bulk viscosity, shear viscosity and  thermal conductivity.  At sufficiently low temperatures phonons give the dominant contribution to the transport coefficients if all the other low energy excitation of the system are gapped.   We first consider  a system  of cold fermionic atoms close to the unitarity  limit.  Then we turn to the superfluid phase of quark matter that may be realized at high baryonic density. 
\end{abstract}
  
\section{Introduction}
 The  condition for superfluidity and  the hydrodynamic equations governing its normal and superfluid components have been derived by  Landau in his pioneering work~\cite{Landau1941}.  Superfluidity is a property of quantum fluids related with  the existence of low energy excitations  that satisfy the Landau's criterion for superfluidity~\cite{Landau1941,IntroSupe,landaustat}
 \begin{equation}\label{landau-criterion}
{\rm Min \frac{\epsilon(p)}{p} \neq 0}\,,
\end{equation}
where $\epsilon(p)$ is the dispersion law of the excitation.

In general, superfluidity is due to the appearance of a  condensate which spontaneously breaks a global  symmetry of the system. As a consequence, in the low energy spectrum one has a Nambu-Goldstone boson, the phonon $\varphi$, with  a linear dispersion law  that satisfies Eq.~(\ref{landau-criterion}). 

Superfluidity was first discovered in $^4$He, which becomes a frictionless fluid  
when cooled at temperatures below 2.17 K~\cite{IntroSupe,landaufluids}. The superfluid property of $^4$He is due to the Bose-Einstein condensation of  the bosonic atoms  in the lowest quantum state; thus  quantum effects become macroscopically observable. 
Fermionic systems can become superfluid as well.  According with the Cooper theorem, fermionic superfluidity takes place in quantum degenerate systems 
when the interaction between neutral fermions is attractive and the temperature is sufficiently low.  In this case one has the formation of a di-fermion condensate that breaks a global continuous symmetry. 

We discuss the hydrodynamic equations  of superfluids  focusing on two  quite interesting fermionic systems. First, we consider   trapped cold atomic gases~\cite{Giorgini:2008zz}  in the region of infinite scattering length (the so-called unitarity limit). Then, we turn to cold relativistic quark matter at extremely high baryonic densities in the color-flavor locked (CFL) phase~\cite{Alford:1998mk}.  These two phases of matter are quite different, however they share an important property: both systems are (approximately) scale  invariant, and their properties  do not depend on the detailed form of the interaction.

The hydrodynamic equations governing the  fluctuations of  a superfluid are essentially different from standard fluid equations. In a superfluid there  are two  independent motions, one normal and the other superfluid. The transport properties depend on the shear viscosity coefficient,  $\eta$, on  three independent bulk viscosity coefficients, $\zeta_1, \zeta_2, \zeta_3$, and on the thermal conductivity $\kappa$. These quantities  can be understood as phenomenological coefficients which relate the rate of change of some quantity with the corresponding affinity~\cite{prigogine}. The requirement that the dissipative processes lead to positive entropy production imposes that 
$\kappa, \eta, \zeta_2, \zeta_3$   are positive and that  $\zeta_ 1^2 \leq \zeta_2 \zeta_3$.    The bulk viscosity  coefficient  $\zeta_2$ plays the role of the standard bulk viscosity coefficient. On the other hand, $\zeta_1$ and $\zeta_3$  provide a coupling between the hydrodynamic equations of the two components.
The friction forces due to bulk viscosities can be understood as drops, with respect to their equilibrium values, in the main driving forces acting on the normal and 
superfluid components. These forces are given by the gradients of the pressure $P$ and of the chemical potential $\mu$. One can write in the comoving frame
\begin{eqnarray}
P &=& P_{\rm eq} - \zeta_1 {\rm div}(V^2 {\bf w}) -\zeta_2  {\rm div} {\bf u}\,, \\
\mu &=& \mu_{\rm eq} - \zeta_3 {\rm div}(V^2 {\bf w}) -\zeta_1  {\rm div}{\bf u}\, ,
\end{eqnarray}
where $P_{\rm eq}$ and $\mu_{\rm eq}$ are the equilibrium pressure and chemical potential, $V$ is a quantity proportional
to the quantum condensate, $\omega^\mu = - \left(\partial^\mu \varphi + \mu u^\mu \right)$ and $u^\mu$ is the  velocity of the fluid.

In a conformally invariant system it has been shown in Ref.~\cite{Son:2005tj} that  $\zeta_1 = \zeta_2 =0$. However, $\zeta_3$, $\kappa$ and $\eta$  cannot be determined by the same symmetry reasoning. Regarding the shear viscosity, it is worth mentioning  that a lower bound for the shear viscosity to entropy ratio has been derived employing the AdS/CFT correspondence in the strong coupling limit within the $N=4$ super-symmetry Yang Mills theory~\cite{Kovtun}, obtaining $\eta/s = 1/4 \pi$.  

In the low temperature regime, $T \ll T_c$, where $T_c$ is the critical temperature for superfluidity, the transport properties of superfluids are determined  by  phonons. The contribution of   other degrees of freedom is thermally suppressed.   In this case one can show that $\zeta_1^2= \zeta_2 \zeta_3$, meaning that there are only two independent bulk viscosity coefficients and that one of the relation for positive entropy production is saturated; the system tends toward the state where the velocity of the superfluid component and the velocity of the normal component are parallel
and bulk viscosity does not lead to dissipation.
 
For $T \ll T_c$ the transport coefficients strongly depend on the phonon dispersion law. 
The shear viscosity is the only transport coefficient that does not vanish for phonons with a linear dispersion law. But, for the bulk viscosities and for the thermal conductivity  one  has to include the term cubic in momentum 
\begin{equation}\label{dispersion}
\epsilon(p)= c_s p + B p^3  + {\cal O}(p^5)\,.
\end{equation}
Moreover, in the computation of the bulk viscosity one has to consider the processes that change the number of phonons. The parameter $B$  determines whether  some processes are or are not kinematically allowed.  For $B>0$ the leading contribution comes from the Beliaev process  $\phi \to \phi \phi$.
In the opposite  case the Beliaev process is not kinematically allowed and one has to consider the processes  
$\phi \phi \to \phi \phi \phi$.

\section{Cold atoms at unitarity}
Experiments with trapped cold atomic gases have reached an extremely high level of accuracy.  The system consists of fermionic atoms, like $^6{\rm Li}$ or  $^{40} {\rm K}$, in two different hyperfine states. The fermions in the two hyperfine states have opposite spin and the interaction between them can be tuned by means of  a magnetic-field Feshbach resonance~\cite{Ohara:2002}. The strength of the interaction between atoms  depends on  the applied  magnetic field and can be  measured  in terms of the $s$-wave scattering length. By varying the magnetic-field controlled interaction,  fermionic pairing is observed to undergo the Bose-Einstein condensate (BEC) to Bardeen-Cooper-Schrieffer (BCS) crossover. In the weak coupling BCS region  the system is characterized by the formation of Cooper pairs. In the  strong coupling limit the system can be described as a  BEC dilute gas. 
The unitary limit is reached when the magnetic field is tuned at the Feshbach resonance~\cite{Feshbach}, where the  two-body scattering length  diverges. 

Far from  unitarity, the properties of the   system  are qualitatively and quantitatively well understood using mean field theory~\cite{BEC}.  However, the mean field expansion is not reliable close to  unitarity because the scattering length is much larger than the inter-particle distance and  there is no small parameter in the Lagrangian to expand in. Therefore  fluctuations may change the mean field results substantially. 

Close to the unitarity region  quantitative understanding of the phases comes  from Monte-Carlo simulations~\cite{Carlson:2005kg}, or considering the expansion in a  small parameter that comes from the generalization to an arbitrary number $N$ of spins~\cite{Radzihovsky:2007}, or considering  an $\epsilon= 4-d$ expansion and then extrapolating the results to $d=3$ dimensions.

A different method comes from considering that  at sufficiently low temperature the only active degrees of freedom are the phonons (see however~\cite{Bulgac:2005pj}). The interesting point is that the effective Lagrangian for phonons can be determined from the pressure and  by demanding  non-relativistic general coordinate invariance and conformal invariance~\cite{Son:2005rv}
\begin{equation}\label{L-cb}
{\cal L}_{\rm eff} =  c_0 m^{3/2} X^{5/2} +  c_1 m^{1/2} \frac{(\nabla X)^2}{\sqrt{X}} + \frac{c_2}{\sqrt{m}}(\nabla^2 \phi)^2  \sqrt{X} \,,
\end{equation}
where $c_0$, $c_1$ and $c_2$ are three dimensionless and universal constants. From the expression above one can determine the coefficients appearing in the  dispersion law of phonons obtaining
\begin{equation}\label{dispersion-conf}
c_s= \sqrt{\frac{2 \mu}{3}} \qquad {\rm and} \qquad B = - \pi^2 c_s \sqrt{2 \xi}\left(c_1+ \frac{3}{2} c_2\right) \frac{1}{k_F^2}\,.\end{equation}
The hydrodynamic equations describing the behavior of the system are
\begin{eqnarray}\label{j-dis}
\frac{\partial j_i}{\partial t}  + \partial_j(\Pi_{ij}+ \tau_{ij}) &=&0 \,,\nonumber\\ \label{v-dis}
\frac{\partial {\bf v_s}}{\partial t} + \nabla \left( \mu + \frac{{\bf v_s}^2}{2}+ h\right) &=& 0 \,,\\ \label{E-dis}
 \frac{\partial E}{\partial t} + \nabla \cdot ({\bf Q} + {\bf Q}^\prime) &=&0 \nonumber\,, 
\end{eqnarray}
where
\begin{equation}
{\bf Q}^\prime =  {\bf q} + h ( {\bf j} - \rho  {\bf v}_{n}) +  \tau \cdot {\bf v}_{n} \,,
\end{equation}
and  $\tau_{ij}$, $h$ and ${\bf q}$ are small dissipative terms that  close to equilibrium are given by 
\begin{eqnarray}\label{tau}
 \tau_{ij} &=& - \eta\big(\partial_j   v_{ni} + \partial_i v_{nj} - \frac{2}{3} \delta_{ij} \nabla \cdot
 {\bf v}_n \big) -\delta_{ij} \big(\zeta_1 \nabla \cdot (\rho_s({\bf v}_s -  {\bf v}_n)) + \zeta_2 \nabla \cdot {\bf v}_n\big) \,, \nonumber\\ \label{h}
 h &=& -\zeta_3\nabla \cdot (\rho_s({\bf v}_s -  {\bf v}_n))- \zeta_1 \nabla \cdot {\bf v_n} \,,\nonumber\\
{\bf q} &=& - \kappa \,\nabla T \nonumber\,.
\end{eqnarray}
The various transport coefficients can be determined starting from the phonon Lagrangian in Eq.~(\ref{L-cb}), considering the appropriate  phonon scattering process. The shear viscosity  of a unitary superfluid at  low temperature  has been computed in Ref.~\cite{Rupak:2007vp} obtaining that   
\begin{equation}
\frac{\eta}{s} \simeq 7.7 \times 10^{-6} \xi^5 \frac{T_F^8}{T^8}\,,
\end{equation}
where $\xi=0.2-0.3$ and $T_F$ is the Fermi temperature.

The thermal conductivity from phonons of a unitary gas has not been determined, yet.

Regarding the bulk viscosity one has to consider phonon number changing processes. For cold atoms at unitarity $B>0$ and the Beliaev process is kinematically allowed.   In Ref.~\cite{Escobedo:2009bh} it has been verified that   in the conformal limit $\zeta_1 = \zeta_2=0$, while  the remaining bulk viscosity coefficient evaluated within kinetic theory in the relaxation time approximation turns out to be
\begin{equation}
\zeta_3  \simeq   3695.4 \left( \frac \xi \mu \right)^{9/2} \frac{(c_1 + \frac 32 c_2)^2}{m^8} T^3 + {\cal O}\left(T^5\right) \,.
\end{equation}

The presence of non-vanishing transport coefficients leads to  experimentally detectable effects. The damping of radial breathing mode depends on the bulk and shear viscosity coefficients. However, since $\zeta_2$ vanishes, the bulk viscosity enters only in presence of a difference of velocity between the normal and the superfluid component and is in general negligible. In order to  determine $\zeta_3$ one should produce oscillations where the normal and superfluid component oscillate out of phase.   The transport coefficients enter also into the damping rate for  the propagation of first and second sound in a superfluid~\cite{IntroSupe}. The damping of first sound, $\alpha_1$,  depends on the shear viscosity  and on $\zeta_2$, whereas the damping of second  sound, $\alpha_2$, depends on all  the dissipative coefficients.

\section{Quark matter at extremely high baryonic density}

Relativistic superfluids
might be realized  in the interior of neutron stars where the temperature is low and the typical energy scale of  particles is extremely high.   In particular, in the inner crust of neutron stars the attractive interaction between neutrons can lead to the formation of a BCS condensate.
Moreover, if deconfined quark matter is present in the core of neutron stars
it will very likely be in a color superconducting phase~\cite{reviews}.
Quantum Chromodynamics (QCD) predicts that
at asymptotically high densities  quark matter is in the color-flavor locked  phase.
In this phase up, down and strange quarks of all three colors pair forming a difermion condensate that is antisymmetric in color and flavor indices.
CFL quark matter is  a superfluid as well, because by picking a phase its order parameter breaks the quark-number $U(1)_B$ symmetry spontaneously.

There are different formulations of the non-dissipative  hydrodynamical equations of  a relativistic superfluid~\cite{khalatnikov,Son:2000ht}. They were derived as
relativistic generalizations of  Landau's two-fluid model of non-relativistic superfluid dynamics. 
The dissipative terms which enter into the relativistic hydrodynamical equations were derived in~\cite{Gusakov:2007px}. As it occurs in the non-relativistic case, for a relativistic superfluid  one can define the thermal conductivity, $\kappa$,
the shear viscosity coefficient, $\eta$, and three independent bulk viscosity coefficients, $\zeta_1,\zeta_2,\zeta_3$.

The dispersion law of phonons in the CFL phase  has been derived in~\cite{Zarembo:2000pj},  and one has 
\begin{equation}
c_s= \sqrt{\frac 1 3} \qquad {\rm and} \qquad B= -\frac{11 c_s}{540 \Delta^2} \,.
\end{equation}
 In the asymptotic high density limit, the gap, $\Delta$, can be computed from QCD 
~\cite{Son:1998uk}
\begin{equation}
\Delta \simeq b_0\mu g^{-5} \exp\left(-\frac{3 \pi^2}{\sqrt{2} g}\right) \ ,
\end{equation}
where $b_0 = 512 \pi^4 (\frac 23)^{5/2} \exp{ \left(-\frac{\pi^2+4}{8} \right)}$ and $g$ is the QCD gauge coupling constant.  At very high chemical potentials, one can neglect  quark masses,  as far as $m_q \ll \mu$;
moreover the coupling constant is small, $g (\mu) \ll 1$, and one can assume that CFL is approximately scale invariant. 

A  first study of the shear viscosity and of the contribution to the bulk viscosity coefficients  due to phonons has been presented in Refs.~\cite{Manuel:2004iv,Manuel:2007pz}. Beside phonons, kaons  may give a sizable contribution to the transport coefficients. The contribution of kaons to $\zeta_2$  has  been studied in Ref.~\cite{Alford:2007rw}.  There is still no computation of the contribution of kaons to the remaining bulk viscosity coefficients. 

Considering only the contribution of phonons one has that $\zeta_1=\zeta_2=0$,  while   the third bulk viscosity coefficient does not vanish and depends parametrically on the physical scales as
\begin{equation}
\zeta_3 \sim \frac{1}{T} \frac{\mu^6}{\Delta^8} \, .
\end{equation}
We remark that 
 these are only  approximated results that arises in the $g \ll 1$ limit, after neglecting the running of the QCD gauge coupling constant and the effect of the strange quark mass. 

The contribution to thermal conductivity due to phonons and kaons has recently been studied in Ref.~\cite{Braby:2009dw}.  The thermal conductivity from phonons turns out to be dominant and given by 
\begin{equation}
\kappa \sim 6 \times 10^{-2} \frac{\mu^8}{\Delta^6}\,,
\end{equation}
whereas the shear viscosity from phonons as determined in~\cite{Manuel:2004iv} is given by 
\begin{equation}
\eta \simeq 1.3 \times 10^{-2} \frac{\mu^8}{T^5} \,.
\end{equation}

If superfluidity occurs in the interior of compact stars, it should be possible to find 
signatures of its presence through a variety of astrophysical phenomena. For example, the most natural 
explanation for the sudden spin-up of pulsars~\cite{Anderson:1975zze}, the so-called glitches, relies on the existence of
a superfluid component in the interior of the star, rotating much faster than the outer solid crust.
After the unpinning of the superfluid vortices, there is a transfer of angular momentum from the interior of the star to the outer crust, giving rise to the the pulsar glitch.

Another possibility to detect or discard the presence of relativistic superfluid phases 
consists in studying the evolution of the r-mode oscillations of  compact stars~\cite{Andersson:2000mf}. R-modes are non-radial oscillations of the star with the Coriolis force acting as the restoring force. They provide a severe limitation on the rotation frequency of the star through coupling to  gravitational radiation (GR) emission. When dissipative phenomena damp these r-modes the star can rotate without losing  angular momentum to GR. If dissipative phenomena are not strong enough,  these  oscillations  
will grow exponentially and the star will keep slowing down until some dissipation mechanism   is able to  damp the r-mode oscillations. Therefore,  the study of r-modes  is useful in  constraining  the stellar structure and can be used to rule out some matter phases.
For such studies it is necessary to consider in detail all the dissipative processes and to compute the corresponding  transport coefficients.

\section{Mutual friction}
Beside shear and bulk viscosity in a rotating superfluid  one has to consider one more dissipative process. This is due to the scattering of phonons off superfluid vortices  and leads to the so-called mutual friction force between the normal and the superfluid components. 
This force can be evaluated  from the differential cross section per unit vortex length for the phonon-vortex scattering process. In case the phonons form a diluted gas, one has that  
\begin{equation}
\frac{d\sigma}{d\theta} = \frac{c_s}{2 \pi E} \frac{\cos^2{\theta}}{\tan^2{\frac\theta 2}} \sin^2{\frac{\pi E}{\Lambda}} \,,
\end{equation}
where $\theta$ is the scattering angle,  $E$ is the phonon energy and   $
1/\Lambda =\left(1- c_s^2\right)\frac{k}{2 \pi c_s^2} $,
with $k$ is  the quantized circulation.

\section{Addendum: Gravity analogs}
The evaluation of the cross section for the phonon-vortex scattering can be determined employing the gravity analogs technique.  According with Unruh~\cite{Unruh:1980cg}, the effective action of a phonon propagating in a fluid is equivalent to the action of a scalar field in a curved background. This analogy can be used in two different ways; to study some aspects of general relativity  (GR), e.g. black hole evaporation, from the analogous hydrodynamical system; to study some properties of hydrodynamics using results of general relativity.  In our case we want to determine the phonon-vortex cross section in the CFL phase, using the analogous results obtained in GR~\cite{Mannarelli:2008jq}. In order to do this  we separate the Nambu-Goldstone boson field as 
\begin{equation}
\varphi (x) = \bar \varphi(x) + \phi(x) 
\end{equation}
where $\bar \varphi(x)$ is a classical field describing the bulk motion of the superfluid and $\phi(x) $ is the fluctuations on the top of the superfluid.
Thus we can write
\begin{equation}
S[\varphi] = S[\bar \varphi] + \frac 12 \int d^4 x \,\frac{ \partial {\cal L}_{\rm eff} }
{\partial(\partial_\mu \varphi) \partial(\partial_\nu \varphi)}
\Bigg \vert_{\bar \varphi}\partial_\mu \,\phi \partial_\nu \phi + \cdots 
\end{equation}
and introducing the acoustic metric 
\begin{equation}
{g}_{\mu \nu} = \eta_{\mu\nu} + (c_s^2-1)v_\mu v_\nu\,,
\end{equation}
we have the action for the phonon field
\begin{equation}S[\phi] = \frac 12 \int d^4 x \sqrt{- { g} } \, { g}^{\mu \nu} \partial_\mu \,\phi \partial_\nu \phi \,.
\end{equation}
When  the classical background has a vortex configuration it results in a non-trivial metric. Then one can obtain the vortex-phonon cross section studying the propagation of the scalar field in this metric~\cite{Sonin:1987zz}.

\noindent\underline{\it Acknowledgment}\ This work was supported by the Spanish grant FPA2007-60275.

%

\end{document}